\documentclass[12pt]{article}
\usepackage{aaspp4,rotating}
\begin{document}
\slugcomment{Accepted for publication in ApJL}
\title{FU Orionis resolved by infrared long baseline interferometry at a
  2-AU scale}
\author{F.~Malbet\altaffilmark{1}, J.-P.~Berger\altaffilmark{1},
  M.M.~Colavita\altaffilmark{2}, C.D.~Koresko\altaffilmark{3},
  C.~Beichman\altaffilmark{4}, A.F.~Boden\altaffilmark{2},
  S.R.~Kulkarni\altaffilmark{3}, B.F.~Lane\altaffilmark{3},
  D.W.~Mobley\altaffilmark{2}, X.P.~Pan\altaffilmark{3},
  M.~Shao\altaffilmark{2}, G.T.~Van Belle\altaffilmark{2},
  \& J.K.~Wallace\altaffilmark{2}}
\altaffiltext{1}{Laboratoire d'Astrophysique, Observatoire de Grenoble, UMR
  UJF/CNRS 5571, BP 53, F-38041 Grenoble cedex 9, France}
\altaffiltext{2}{Jet Propulsion Laboratory, California Institute for
  Technology, Pasadena, CA} 
\altaffiltext{3}{Palomar Observatory, California Institute for Technology,
  Pasadena, CA} 
\altaffiltext{4}{Infrared Processing and Analysis Center, California
  Institute for Technology, Pasadena, CA}

\begin{abstract} 
  We present the first infrared interferometric observations of a young
  stellar object with a spatial projected resolution better than 2~AU. The
  observations were obtained with the {\em Palomar Testbed Interferometer}.
  FU~Ori exhibits a visibility of $V^2=0.72\pm0.07$ for a $103\pm5$ m
  projected baseline at $\lambda=2.2\micron$.  The data are consistent on
  the spatial scale probed by PTI both with a binary system scenario
  (maximum magnitude difference of $2.7 \pm 0.5\mbox{ mag}$ and smallest
  separation of $0.35 \pm 0.05\mbox{ AU}$) and a standard luminous
  accretion disk model ($\dot{M}\sim 6\times10^{-5}\mbox{ M$_\odot$
    $yr^{-1}$}$) where the thermal emission dominates the stellar
  scattering, and inconsistent with a single stellar photosphere.
\end{abstract}

\keywords{stars: pre-main sequence --- circumstellar matter --- stars:
  individual (FU Ori) --- accretion disks --- instrumentation:
  interferometers --- infrared: stars}

\section{Introduction}
\label{sect:intro}

FU Orionis is the prototype of a class of young stellar objects (YSOs,) 
called {\em FUors,} that have undergone photometric outbursts of the order 
of 4-6 magnitudes in less than one year (Herbig 1966).  A FUor's luminosity 
typically peaks at $\sim 500 \mbox{ L$_\odot$}$, and then appears to decay 
on a 100-year timescale.  FUors exhibit large infrared excesses, 
double-peaked line profiles, apparent spectral types that vary with 
wavelength, broad, blueshifted Balmer line absorption, and are often 
associated with strong mass-outflows (see Hartmann \& Kenyon 1996 for a 
recent review on this phenomenon).

The FUors have been convincingly modeled as low-mass pre-main sequence 
stars (T Tauri stars) which are surrounded by luminous accretion disks.  
The inferred peak accretion rates are on the order of $10^{-4}\mbox{ 
M$_\odot$ yr$^{-1}$}$.  The energy released by the accretion process is 
radiated at the disk surface, overwhelming the stellar emission.  Kenyon \& 
Hartmann (1991) showed that an opaque, dusty $A_v \sim 50 \mbox{ mag}$ 
infalling envelope with a cavity along our line of sight is consistent with 
both the mid-IR excess of FUors and the relatively low $A_v \sim 
1-4\mbox{ mag}$ estimated extinction of the inner source.

The radiative balance between disk emission and gravitational energy released
by the accretion process implies that the disk's temperature falls with the
$-3/4$ power of radius, leading to anticipated angular sizes on the order of
one milliarcsecond in the near-IR at a distance of 450 pc.  Malbet \& Bertout
(1995, hereafter MB95) therefore proposed using infrared long-baseline
interferometry to probe the physics of these disks.

We present in this paper the first infrared interferometric observations of 
a YSO, taken with 4 milliarcsecond resolution at 2.2 $\mu$m using the {\em 
Palomar Testbed Interferometer (PTI)}.  The corresponding linear resolution 
is better than 2~AU at the 450~pc distance to FU Orionis.  The object is 
found to be clearly resolved, with a fringe visibility significantly below 
unity and consistent with, although not unique to, the predictions of the 
disk model.  Section \ref{sect:obs} presents the observations and Section 
\ref{sect:proc} the data processing.  In Section \ref{sect:interp}, we 
discuss the results in the context of a number of models.

\section{Observations}
\label{sect:obs}

FU Orionis was observed from 1997 November 1 to 7 (nights 305, 306, 307, 
309, 310, and 311) using the PTI.  The PTI, located adjacent to the 
200-inch Hale telescope on Palomar Mountain (California, USA), is a 
two-element interferometer with a 110-m baseline oriented roughly 
North-South.  The observations were taken in the ``single-star" mode, in 
which the measurement consists of the squared fringe visibility $V^2$ along 
the single baseline (Colavita et al.\ 1994).  The fringe was tracked and 
measured in the K band for 11 scans of 125-130 sec each.  Earth rotation 
provided a limited range of projected baselines: $98\mbox{ m}\leq B \leq 
108\mbox{ m}$ and $55^{\circ}\leq \theta \leq 73^{\circ}$. We used HD 
42807 located at $9.4^{\circ}$ from FU Ori as local calibrator 
star\footnote{Characteristics from Hipparcos catalogue (Perryman et al.\ 
1997): $\mbox{V}=6.5$, G8V, 55 mas parallax.  Estimated K magnitude: 4.7; 
estimated diameter: $0.45\pm0.03\mbox{ mas}$.}.  Scans of FU Ori were 
alternated with measurements of HD 42807 to ensure an accurate 
determination of the system visibility.  We also used the other calibrator 
stars from the night to estimate the system visibility (see Sect.
\ref{sect:proc}).

The apparent magnitudes for FU~Orionis are $\mbox{V}=8.9$,
$\mbox{R}=7.7$ and $\mbox{K}=4.6$ mag.  The object is close to the flux limit
of both the acquisition and angle-tracking system and the fringe tracker, so
special care was taken during the observations and data reduction to optimize
sensitivity and avoid biasing the results.  The instrument observing
parameters were optimized for FU Ori, and identical settings were used for HD
42807.  At the level of accuracy of this experiment, the global calibration
is consistent with that obtained with the local calibrators.

\section{Data processing}
\label{sect:proc}

Because the faintness of FU~Orionis made it necessary to operate in
previously little-explored regimes close to the sensitivity limits of the
instrument, we present the data reduction procedures and reliability tests
here in some detail.  The data processing essentially followed the steps
detailed in Colavita (1998).  We base our results on the spectrometer
data rather than the white light data, since the high visibilities produced
by its single-mode fiber and narrow bandpass outweighed the larger photon
rate in the broadband ``white-light" channel.  To avoid introducing biases
and to maximize sensitivity, we employed an incoherent estimator averaged
over the entire K band.  Examination of the other data products was used to
help confirm the robustness of the final visibility measurement.

\subsection{Calibrated visibilities}
\label{sect:proc.calib}

We obtained the {\it raw} square visibilities $V^2_{\rm raw}\sim 0.6\pm0.05$
for FU Ori.  The critical next step in the processing consists of dividing
the raw visibilities by an estimation of the instrument + atmosphere
visibility ({\it i.e.,} the system visibility) to obtain {\it calibrated} 
visibilities $V^2$.

The single-mode fiber eliminates most atmospheric effects except for {\it
fringe jitter,} which is due to differential atmospheric piston.  The fringe
jitter can introduce a bias into the measured visibility which must be
estimated and corrected.  The first-difference variance phase
$\sigma_{\Delta\phi}$ gives an estimate of the jitter which can be used to
derive a small multiplicative correction (Colavita 1998).  The jitter
correction for these FU~Orionis data ranges from $0.94\pm0.05$ to
$0.98\pm0.03$.

The calibrator stars are used to estimate the system visibility.  The
Hipparcos catalogue (Perryman et al.\ 1997) provides the spectral type and
parallax for each calibrator, from which we estimated their angular
diameters.  The calibrators are assumed to have uniform surface brightness.
The results are not sensitive to this assumption since the calibrator stars
are chosen to have diameters much smaller than the interferometer
resolution.  The system visibilities are fairly constant for each night (see
Section \ref{sect:proc.qual}) showing that the instrument is rather stable, 
with $V^2_{\rm inst}\sim 0.87\pm0.02$.

\begin{figure}[p]  
  \begin{center}
    \leavevmode
    \includegraphics[width=0.6\textwidth]{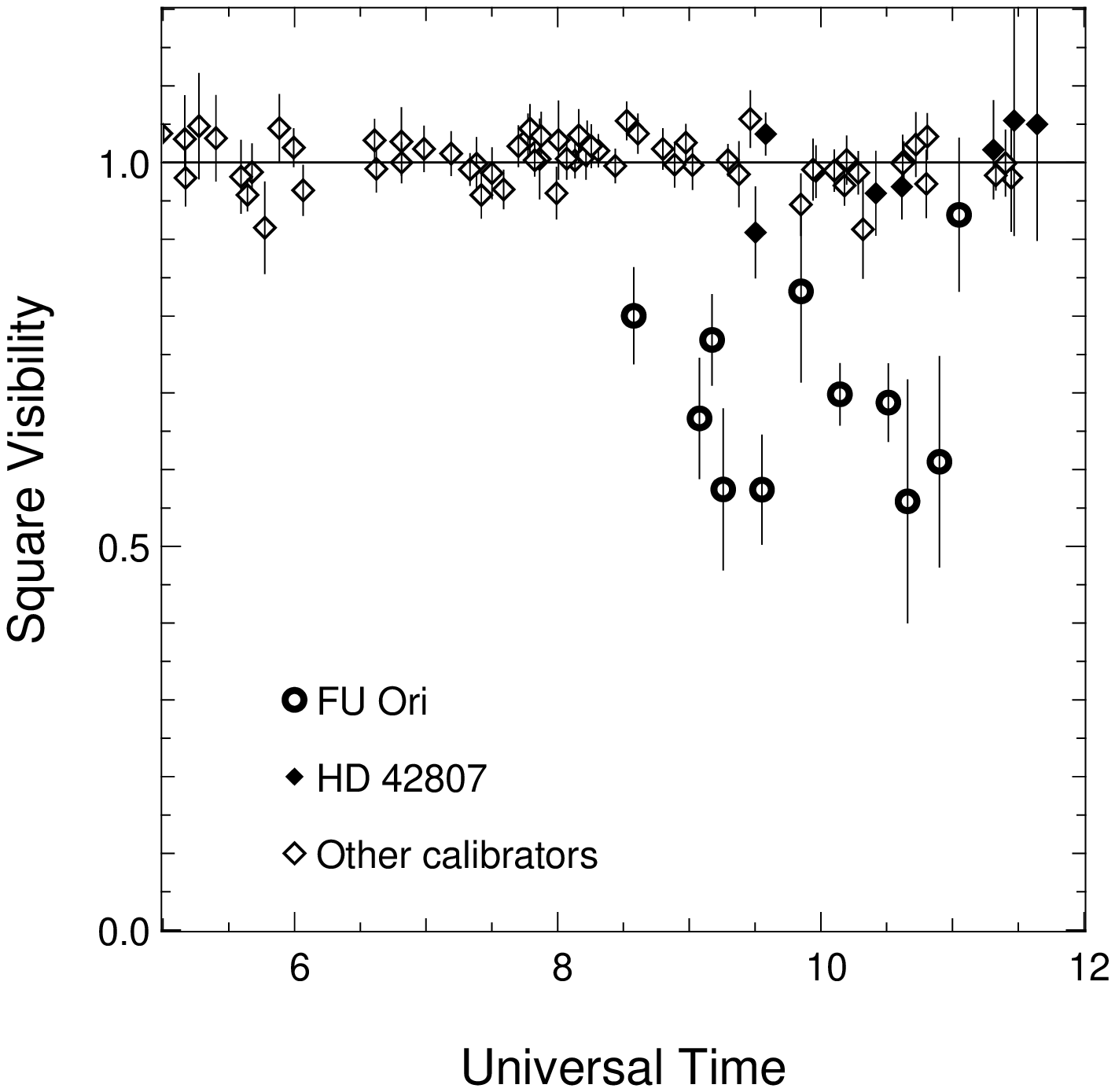}
    \caption{Calibrated square visibilities of FU Ori (open circles) and
      its calibrators (diamonds) from several nights. The local calibrator
      HD 42807 is displayed in filled diamonds.\label{fig:results}}
  \end{center}
\end{figure}

Figure \ref{fig:results} displays the calibrated square visibilities $V^2$
for FU Orionis and its calibrators.  The error bars are estimated from the
contributions of the fluctuations of $V^2_{\rm raw}$ between subsamples of
the individual scans, the jitter correction errors, and the instrumental
visibility errors.

\subsection{Data Quality Measures}
\label{sect:proc.qual}

\begin{figure}[p]
  \begin{center}
    \leavevmode
    \includegraphics[angle=-90,width=\textwidth]{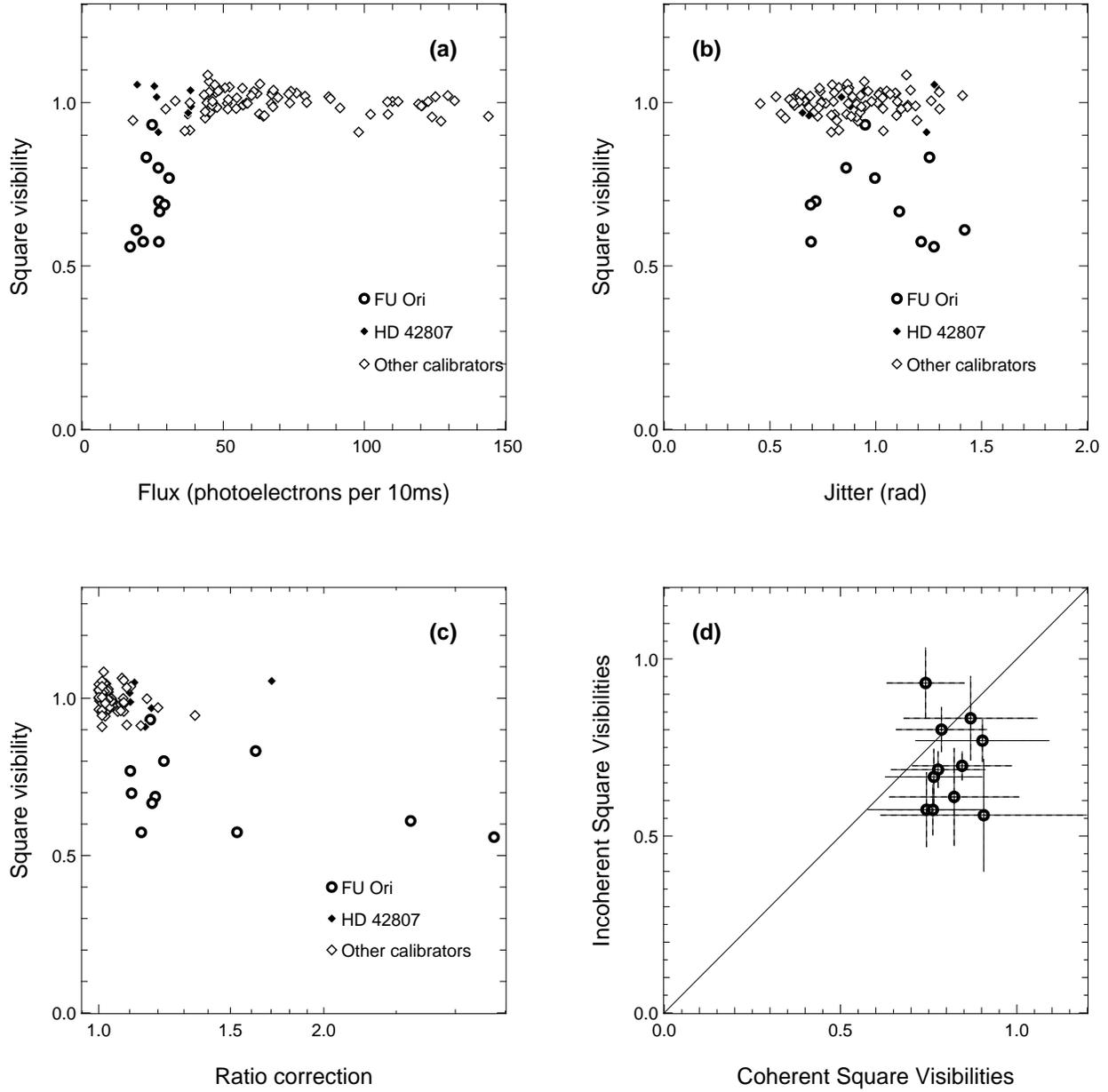}
    \caption{Data quality measures. Square visibilities versus flux in
    units of photoelectrons per 10 ms readout (Panel a), versus jitter
    (Panel b), versus ratio correction (Panel c). Incoherent (y-axis)
    versus coherent (x-axis) square visibilities (Panel d); the crosses
    represent the error bars.
    \label{fig:quality}}
  \end{center}
\end{figure}

Because the magnitude of FU Ori is close to the limiting magnitude of the 
PTI, we present the following unusually detailed discussion of the data
quality.  We have run several checks on the data to validate the results and
the associated uncertainties.  The corresponding plots are displayed in Fig.\
\ref{fig:quality} and discussed below:

{\bf Visibility vs Stellar Flux.} One might imagine that for an object at
the detection limit, flux-dependent terms might become important in the
calibration of the visibilities.  Figure\ \ref{fig:quality}a displays $V^2$
data for FU Orionis and its calibrators and shows no decrease in visibility
as a function of flux.

{\bf Read-out Noise.} The observations of FU Ori are dominated by the 
detector read-out noise.  Colavita (1998) assesses the expected 
uncertainties for PTI data.  In our observations, the flux $N\sim20$ 
photoelectrons per 10ms read-out, the read noise $\sigma\sim16\mbox{ 
e$^-$}$ and the total number of samples $M=12500$ (5 spectral channels 
combined incoherently each with $25\mbox{ s}$ of 10-ms frames) lead to 
$\sigma_{V^2}=0.113$ for each 25-s measurements.  The observed statistical 
error of $0.05$ for a 125-s long observation is consistent with the errors 
computed statistically for $V^2_{\rm raw}$.

{\bf Phase Jitter}.  The effect of the phase jitter due to imperfect tracking
of the fringe as it moves due to atmospheric turbulence (piston) is to
introduce a bias which decreases the estimated visibility.  Colavita
(1998) shows that the jitter is related to the convolution of the variance,
$\sigma^2_{\Delta\phi}$, of the differences between successive phase
measurements with the power spectrum of the piston disturbance.  If the power
spectrum is dominated by frequencies higher than 10 ms, one finds that the
visibility bias is $e^{ - C_{\Gamma} \sigma^2_{\Delta\phi}}$ with
$C_{\Gamma}=0.04$.  A plot of the jitter-corrected $V^2$ versus
$\sigma_{\Delta\phi}$ (Fig.\ \ref{fig:quality}b) shows no systematic
dependence of the visibility on the size of the jitter.  We conclude that the
data are free of jitter bias.  The errors introduced by this jitter
correction are of the order of $0.05$ for nights 305 and 306, $0.03$ for
night 307 and $0.04$ for nights 309 to 311.

{\bf Flux Ratio.} The flux-ratio correction accounts for the difference in
fluxes between the two arms of the interferometer, which occurs mainly as a
result of optical vignetting.  This vignetting results in a bias that
decreases the visibility estimation.  If this effect is large, measured
visibilities should decrease as the correction ratio increases.  Figure\
\ref{fig:quality}c displays the flux-ratio correction versus $V^2$.  Except
for the points measured at 10:54 on night 305 and at 10:39 on night 306 that
show strong departures, we see no flux-correction effect in the data and have
made no correction for it.

{\bf Incoherent Data vs Coherent Data.} We processed the coherent data in 
the same manner as the incoherent data.  The effect of jitter is much more 
important in these data\footnote{$C_{\Gamma}\sim 0.4\pm0.2$, measured by 
fitting $K\exp( -C_{\Gamma} \sigma^2_{\Delta\phi})$ to the data.}, giving
rise to larger uncertainties.  The two different measurements are consistent
within the uncertainties (Fig.\ \ref{fig:quality}d), with the coherent
visibilities being slightly smaller ($V^2\sim0.8$).

{\bf White-Light Data.} The white light channel is not spatially filtered 
by a single-mode fiber.  This leads to low visibilities, $V^2\sim 0.3-0.4$,
and biases that are difficult to understand.  We did not use these
measurements to estimate FU Orionis visibilities.  However, the FU Orionis
raw square visibilities in white light measurements are always smaller than
the calibrator values with a ratio of the order of 0.6-0.7, consistent with
the spectrometer data.

\section{Results and interpretation}
\label{sect:interp}

The calibrated visibilities of FU Orionis display no clear trends with time,
projected baseline, or projected angle.  We have therefore adopted an
averaged visibility value\footnote{Obtained without the two aberrated points
mentioned in \S \ref{sect:proc.qual}.  With all points, the value is
$V^2=0.71\pm0.08$.} of $V^2=0.72\pm0.07$ for an averaged projected baseline
of $103\pm5\mbox{ m}$ and projected angle of $64\pm9^{\circ}$.  Because of
the very limited coverage of the visibility space, the current observations
cannot provide images of the FU Orionis system which would reveal its
morphological structure.  The PTI at $\lambda = 2.2 \micron$ has a spatial
resolution of $\lambda/B=4.4\mbox{ mas}$, corresponding to 2 AU at a distance
of 450 pc.  {\it The principal result of this paper is that FU Orionis is
clearly resolved on this angular scale.}

As described in the introduction, the double-peaked absorption line
profiles, wavelength-dependent spectral types, and excess infrared emission
offer fairly strong evidence for the presence of an active accretion disk.
Although there is strong circumstantial evidence for the disk scenario,
there has not yet been a direct detection of a physical structure of the
predicted extent and we therefore briefly discuss several alternative
interpretations of the new observational result.

{\bf Resolved stellar surface.}
The angular diameter of a uniformly emitting stellar surface that would 
give the observed visibility is 1.55 mas, corresponding to a linear 
diameter of 0.7 AU or $150\mbox{ R$_\odot$}$ at the distance of FU Orionis.  
With a $500\mbox{ L$_\odot$}$ luminosity, such a star would exhibit an 
effective temperature of $2200\mbox{ K}$.  This interpretation is 
inconsistent with the observed spectral type by Kenyon, Hartmann, \& Hewett 
(1988, hereafter KHH88), who find spectral types ranging from F7~I for the 
CN $\lambda 3860$ lines to K3 for the TiO $\lambda 7050$ bands.  We 
therefore consider it unlikely that the PTI observations are resolving a 
bare stellar photosphere.

\begin{figure}[p]
  \begin{center}
    \leavevmode
    \includegraphics[width=0.6\textwidth]{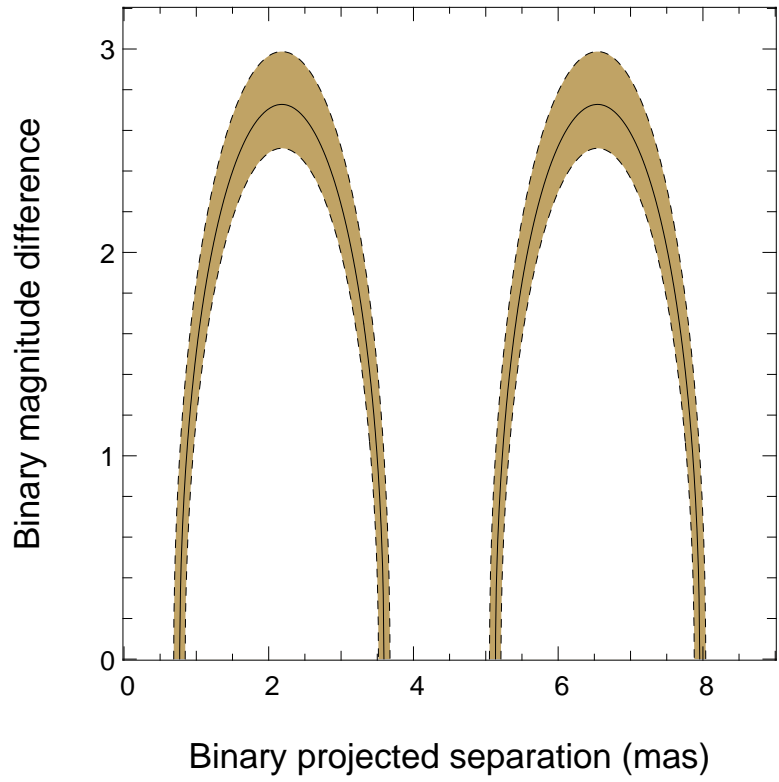}    
    \caption{Binary scenario. The K magnitude difference versus the separation
      projected on the baseline. The colored region is permitted.
      \label{fig:binary}}
  \end{center}
\end{figure}

{\bf Resolved binary system.}
A second phenomenon that could explain the observed visibility is the 
presence of a stellar companion.  Here we consider a model in which FU 
Orionis consists of a pair of unresolved stars with angular separation $s$ 
and magnitude difference $\Delta K$.  Figure \ref{fig:binary} shows the 
values for the binary parameters permitted by our visibility measurement.  
The maximum value of $\Delta K$ consistent with the observations is found 
to be $2.7\pm0.5\mbox{ mag}$, and the smallest separation $s$ is 
$0.8\pm0.1\mbox{ mas}$, i.e.\ $0.35\pm0.05\mbox{ AU}$ at the distance of FU 
Ori.

High resolution spectroscopy (Hartmann \& Kenyon 1985, hereafter HK85) 
reveals double-peaked photospheric lines in FU Ori spectrum that, at least 
in principle, could originate from a binary system with an estimated 
separation $s \leq 0.25 (M \sin^2 i / 1\mbox{ M$_\odot$}) \mbox{ AU}$.  
Such a model is also compatible with our present data set if the total mass 
of the system is $\geq 1.4\mbox{ M$\odot$}$, although a considerably more 
extensive set of observations covering a wide range of projected baseline 
angles could rule it out.  We note that HK85 and KHH88 have favored the 
accretion disk model to explain the line doubling, because it also succeeds 
in modeling the SED and the change of spectral type and line profiles with 
the wavelength.

{\bf Dust halo model.}
Many YSOs show large amounts of scattered light in visible and near-IR images
(e.g.  Nakajima \& Golimowski 1995; Burrows et al.\ 1996).  DeWarf \& Dyck
(1993) fitted their 2.2 $\mu$m speckle observations of FU Orionis with a
Gaussian halo some 0.07 arcsec (35 AU) in diameter, although their data were
also consistent with an unresolved object.  The PTI data can be fitted by an
unresolved star surrounded by a scattering envelope of uniform brightness.
The envelope would account for $\sim 15$\% of the total 2.2 $\mu$m flux.  Its
size is poorly constrained by the PTI visibility measurement: the visibility 
is insensitive to the size if it is larger than $\sim 10 \mbox{ mas}$.
We argue below that depending on the size of the halo, the dust may be seen
by direct radiation rather than scattered light.

The equilibrium temperature of dust close to FU Ori is roughly $270 \mbox{ 
K} (L/1 \mbox{ L$_\odot$})^{1/4} (r/1\mbox{ AU})^{-1/2 {\rm~to~} -3/4}$, 
depending on the geometrical arrangement of the dust and heating source 
(Friedjung 1985).  The temperature of dust 2 AU from this $L \sim 500 
\mbox{ L$_\odot$}$ object, i.e.\ at the minimum physical distance resolved 
by these observations, is $\sim 750$ K; at the 17 AU radius of the halo 
suggested by the speckle results it is $\sim 310$ K.  Hot material close to 
the star would emit strongly at $2.2 \micron$ and with an optical depth 
roughly 5 times greater than the scattering optical depth (Draine \& Lee 
1984).  The ratio of emitted light to scattered light is roughly:
\begin{displaymath}
\frac{B_\nu(T_{\rm gr}(r))}{\omega B_\nu(T_{\rm eff}) (r/R)^{-2}}
\end{displaymath}
where $\omega$ is the albedo ($\sim 0.2$, Draine \& Lee 1984), $r$ is the 
distance to center, $R$ and $T_{\rm eff}$ are the equivalent radius and 
effective temperature of the central source.  With $R=4\mbox{ R$_\odot$}$ 
and $T_{\rm eff}=6000-8000$K, the ratio is much larger than 1 at 2 AU and 
much smaller than 1 at 17 AU.  While a detailed radiative transfer model 
must be used to assess the relative importance of scattering and thermal 
emission, we regard thermal emission as likely to be dominant if the radius 
of the putative dust halo is much smaller than $\sim 10$ AU, whereas 
scattering will be important if the radius is much larger.

\begin{sidewaysfigure}[p]
  \begin{center}
    \leavevmode
    \includegraphics[angle=-90,width=\textwidth]{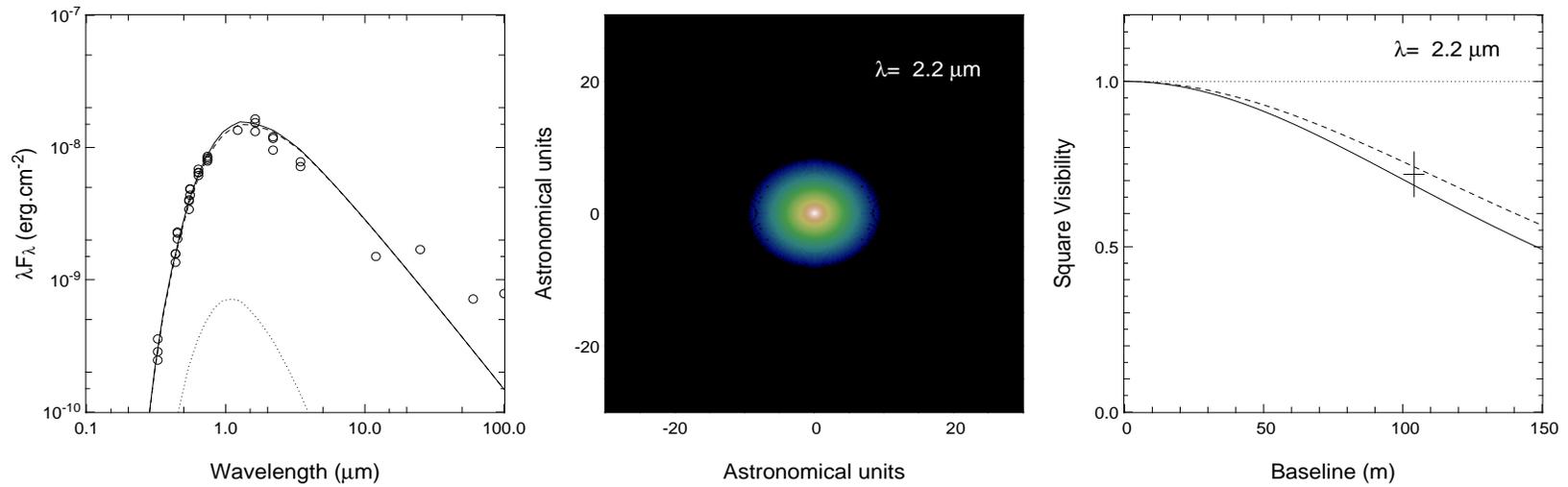}
    \caption{Accretion disk model. Left panel displays the spectral energy
      distribution from litterature data (circles), of the accretion disk
      model (dashed line), the star (dotted line) and the whole system
      (solid line), middle panel the synthetic image of the accretion disk
      at 2.2\micron, and right panel the visibility curves of the accretion
      disk model for the x and y directions (respectively solid and dashed
      lines). The result of our PTI observation of FU Ori is placed on the
      figure with its error bars.
      \label{fig:model}}
  \end{center}
\end{sidewaysfigure}

{\bf Accretion disk model.}
Following MB95, thermal emission from an accretion disk of the type proposed by KH85 and
KHH88 is expected to be resolved in the PTI data, with approximately the
observed fringe visibility. We computed a disk model with a 
surface temperature distribution proportional to $r^{-3/4}$ to fit the
observed SED.\footnote{Photometry data from Allen (1973), Glass \& Penston
(1974), KHH88 and IRAS.} The model implies an accretion rate of
$\dot{M}\sim 6\times10^{-5}\mbox{ M$_\odot$ $yr^{-1}$}$ for a $1\mbox{
M$_\odot$}$ star, an $A_V\sim1\mbox{ mag}$ and an inclination angle of
$i\sim30^{\circ}$.  The resulting synthetic image at 2.2\micron\ is displayed
in Figure\ \ref{fig:model} (middle panel) together with the predicted
visibility curves for the major and minor axis (right panel).  Our
interferometric data are in very good agreement with the accretion disk
model.  However, the precision of the individual visibility measurements is
inadequate to constrain the position angle for the disk.

\section{Conclusions}
\label{sect:concl}

We have resolved a young stellar object for the first time using
long-baseline interferometry in the near-infrared, achieving a projected
spatial resolution of 2 AU using the {\em Palomar Testbed Interferometer}.
Although the single visibility measurement presented here can only offer
limited constraints on existing astrophysical models, it is reassuringly
consistent with the accretion disk which was inferred from earlier spectral
and spectrophotometric data.  More sensitive multi-aperture infrared
interferometers like the {\em Keck Interferometer} and the {\em Very Large
Telescope Interferometer} (VLTI), now under construction, will soon enable
more robust studies by producing true images of FUor disks with $\sim 2$ AU
resolution, and even more detailed images of the disks of less luminous T
Tauri stars.

\newpage
%\begin{center}
%  {\bf Figure captions}\\*[2em]
%\end{center}
%\figcaption{Calibrated square visibilities of FU Ori (open circles) and
%    its calibrators (diamonds) from several nights. The local calibrator HD
%    42807 is displayed in filled diamonds.
%    \label{fig:results}}  

%\figcaption{Data quality measures. Square visibilities versus flux in
%    units of photoelectrons per 10 ms readout (Panel a), versus jitter
%    (Panel b), versus ratio correction (Panel c). Incoherent (y-axis)
%    versus coherent (x-axis) square visibilities (Panel d); the crosses
%    represent the error bars.
%    \label{fig:quality}}

%\figcaption{Binary scenario. The K magnitude difference versus the separation
%    projected on the baseline. The colored region is permitted.
%    \label{fig:binary}} 

%\figcaption{Accretion disk model. Left panel displays the spectral energy
%      distribution from litterature data (circles), of the accretion disk
%      model (dashed line), the star (dotted line) and the whole system
%      (solid line), middle panel the synthetic image of the accretion disk
%      at 2.2\micron, and right panel the visibility curves of the accretion
%      disk model for the x and y directions (respectively solid and dashed
%      lines). The result of our PTI observation of FU Ori is placed on the
%      figure with its error bars.
%    \label{fig:model}}

\end{document}